\documentclass[aip, jcp, amsmath, amsthm, amssymb, citeautoscript, reprint]{revtex4-2}
\usepackage{graphicx, float, xcolor, pgf}
\usepackage{physics, siunitx}
\usepackage[caption=false, justification=centerlast]{subfig}

\usepackage{hyperref}
\hypersetup{
    pdfauthor={Amartya Bose},
    pdftitle={Path integral Lindblad master equation through transfer tensor method},
    colorlinks=true,
}

\allowdisplaybreaks

\begin{document}
\title{Path integral Lindblad master equation through transfer tensor method \& the generalized quantum master equation}
\author{Amartya Bose}
\affiliation{Department of Chemical Sciences, Tata Institute of Fundamental Research, Mumbai 400005, India}
\email{amartya.bose@tifr.res.in}
\begin{abstract}
    Path integrals have, over the years, proven to be an extremely versatile
    tool for simulating the dynamics of open quantum systems. The initial
    limitations of applicability of these methods in terms of the size of the
    system has steadily been overcome through various developments, making
    numerical explorations of large systems a more-or-less regular feature.
    However, these simulations necessitate a detailed description of the
    system-environment interaction through accurate spectral densities, which
    are often difficult to obtain. Additionally, for several processes, such as
    spontaneous emission, one only has access to a rough estimation of an
    empirical timescale, and it is not possible to really define a proper
    spectral density at all. In this communication, an approach of incorporating
    such processes within an exact path integral description of other
    dissipative modes is developed through the Nakajima-Zwanzig master
    equations. This method will allow for a numerically exact non-perturbative
    inclusion of the degrees of freedom that are properly described by a bath
    using path integrals, while incorporating the empirical time scale through
    the Lindblad master equation. The cost of this approach is dominated by the
    cost of the path integral method used, and the impact of the Lindbladian
    terms is effectively obtained for free. This path integral Lindblad
    dynamics method is demonstrated with the example of electronic excitation
    transfer in a 4-site model of the Fenna--Matthews--Olson complex with the
    exciton has a propensity of being ``lost'' to the charge transfer state at
    the third chromophore. The impact of different time-scales of abstraction of
    the exciton is illustrated at no extra cost.

\end{abstract}
\maketitle
\emph{Introduction.} Simulation of dynamics of open quantum systems is the holy
grail of Chemistry. The exponential scaling of quantum mechanics with the
number of degrees of freedom combined with the difficulty of describing thermal
dynamics contributes to the unparalleled challenges of simulating the dynamics
of bulk systems. Methods based on propagating wave functions such as density
matrix renormalization group (DMRG)~\cite{whiteDensityMatrixFormulation1992,
    schollwockDensitymatrixRenormalizationGroup2005,
    chanDensityMatrixRenormalization2011}, the family of multiconfiguration
time-dependent Hartree method
(MCTDH)~\cite{meyerMulticonfigurationalTimedependentHartree1990,
    wangMultilayerFormulationMulticonfiguration2003}, and Gaussian wavepacket
dynamics~\cite{hellerTimeDependentApproach1975,
    hellerTimeDependentVariational1976, begusicOntheflyInitioSemiclassical2018},
cannot handle a continuum of vibrational modes or solvent degrees of freedom
held at a constant temperature.

Methods based on reduced density matrices (RDMs) can overcome many of these
difficulties. After judiciously separating the problem into a ``system'' of
interest and the ``environment,'' these methods aim at solving the dynamics
corresponding to the system while incorporating the effects of the environment.
While this system-solvent separation is critical in decreasing the
dimensionality of the system space, it makes the dynamics non-Markovian.
Approximate simulations of RDMs can be done quite simply using the perturbative
Bloch-Redfield master equation~\cite{blochGeneralizedTheoryRelaxation1957,
    redfieldTheoryRelaxationProcesses1957} or through the empirical Lindblad
master equations~\cite{lindbladGeneratorsQuantumDynamical1976,
    goriniCompletelyPositiveDynamical2008}. However, the results obtained from
such simulations are not systematically improvable. Most crucially,
simulations involving these two methods usually assume that the dynamics can
be approximated in a Markovian fashion. These shortcomings are alleviated
for certain class of problems using numerically exact methods for simulating
the dynamics. The hierarchical equations of motion
(HEOM)~\cite{tanimuraTimeEvolutionQuantum1989,
    tanimuraNumericallyExactApproach2020, xuTamingQuantumNoise2022} and the
quasi-adiabatic propagator path integral method
(QuAPI)~\cite{makriTensorPropagatorIterativeI1995,
    makriTensorPropagatorIterativeII1995} based on the Feynman-Vernon influence
functional~\cite{feynmanTheoryGeneralQuantum1963} are the most common
mathematically rigorous frameworks available.

Over the years, a lot of work has gone into improving the performance of these
methods. The original HEOM was primarily limited to handling Drude-Lorentz
spectral densities. Various approaches are now being developed to handle other
spectral densities, including approaches involving Chebyshev
polynomials~\cite{popescuChebyshevExpansionApplied2016,
    rahmanChebyshevHierarchicalEquations2019} and improvements using tensor
networks~\cite{xuTamingQuantumNoise2022}. Similarly, QuAPI has been developed in
a variety of manners to improve the performance. Notable among these are  the
small matrix decomposition of path integral
(SMatPI)~\cite{makriSmallMatrixDisentanglement2020} which is conceptually
similar to the transfer tensor method (TTM) based on dynamical maps for
non-Markovian processes~\cite{cerrilloNonMarkovianDynamicalMaps2014,
    rosenbachEfficientSimulationNonMarkovian2016,
    buserInitialSystemenvironmentCorrelations2017}, and methods based on tensor
networks~\cite{strathearnEfficientNonMarkovianQuantum2018,
    jorgensenExploitingCausalTensor2019, bosePairwiseConnectedTensor2022,
    boseTensorNetworkRepresentation2021}. Tensor network algorithms have also
been extended to handle extended open quantum
systems~\cite{boseMultisiteDecompositionTensor2022} and simulation of
thermal correlation functions~\cite{boseQuantumCorrelationFunctions2023}.
All these developments have enabled simulations of larger systems up to
longer times~\cite{kunduB800toB850RelaxationExcitation2022,
    boseTensorNetworkPath2022, boseImpactSpatialInhomogeneity2023}.

Though lucrative and now significantly more affordable, application of these
rigorous methods require detailed analysis of the system-environment
interactions resulting in calculation of spectral densities. This may not always
be possible. For instance, spontaneous emission of excited states of molecules
in systems with Frenkel exciton transport, or polaritonic dynamics in presence
of leaky cavities are not easily characterized by a spectral density. These
phenomena can be easily handled using the empirical Lindblad master
equations~\cite{lindbladGeneratorsQuantumDynamical1976,
    goriniCompletelyPositiveDynamical2008}, which is not compatible with the path
integral-based methods. One approach to incorporating everything might be to map
the well-characterized environment onto other Lindblad operators, and then solve
the Lindblad master equation for everything. However, while this treats
everything on the same footing, this approach has the disadvantage of making the
dynamics approximate. We are no longer even accounting for the thermal
environment in a numerically exact manner. Implicit in the Lindblad treatment
are the Markovian approximation to the dynamics and the assumption of a weak
system-environment coupling. The other complementary approach may be to try to
postulate a bath and a spectral density for the processes that we don't have the
details for. If one does that, there is a lot of guesswork associated with the
forms of the spectral density and the nature of the bath.  These questions are
not typically well-settled. So, we effectively not only increase the complexity
of the simulation by adding completely fictitious baths to describe the
processes that are not simple to parameterize, but also add extra artefacts
brought in by our choices of these baths.

The ideal approach, therefore, is neither of the two discussed above, but to
resort to a combination of both. We want to describe the open quantum system in
a fully non-Markovian and non-perturbative manner, while incorporating the
Lindblad master equation into it without any additional cost. This work is aimed
at this fundamental goal. This hybrid approach has the advantage of being (1)
ideal in terms of computational costs, since solving the Lindblad master
equation is free compared to the full path integral simulation; and (2) giving
the optimally accurate dynamics according to our ability to characterize the
processes. A similar motivation has led to the development of a recent classical
trajectory-based method by ~\citet{mondalQuantumDynamicsSimulations2023} that
incorporates the Lindblad loss into the partial linearized density matrix
dynamics~\cite{provazzaSemiclassicalPathIntegral2018,
huoCommunicationPartialLinearized2011}. The path integral Lindblad dynamics
developed here is independent of the method of simulation of the open quantum
dynamics, and allows for a decoupling of the path integral simulations for
incoporation of the environment degrees of freedom from the solution to the
Lindblad operators. We develop this path integral Lindblad master equation
approach by going through the Nakajima-Zwanzig master
equation~\cite{nakajimaQuantumTheoryTransport1958,
zwanzigEnsembleMethodTheory1960}, via the transfer tensor
method~\cite{cerrilloNonMarkovianDynamicalMaps2014}. This decoupling has the
added advantage of making the addition of different Lindblad jump operators to a
system completely free.

\emph{Method.} Consider an open quantum system described by the Hamiltonian
\begin{align}
    H & = H_0 + H_\text{env}\label{eq:oqs}
\end{align}
where $H_0$ is the system Hamiltonian and $H_\text{env}$ is the
system-environment Hamiltonian. Also suppose that while most of the environment
has been properly described in $H_\text{env}$, there are features like spontaneous
decay, or leakage of quantum particles, etc. that cannot be characterized
accurately using thermal baths. For these processes, additionally, the system is
also subject to interactions parameterized by a set of Lindblad jump operators
$L_i$. We want to understand the dynamics of the system under the open quantum
Hamiltonian, Eq.~\ref{eq:oqs} and under the action of the Lindblad jump
operators.

If the initial state can be written in a separable form $\rho_\text{total}(0) = \rho(0)
    \otimes \frac{e^{-\beta H_\text{env}}}{Z}$, the time evolution of the reduced density
matrix corresponding to the system in absence of the Lindblad jump operators can
be simulated using path integrals as~\cite{makriTensorPropagatorIterativeI1995,
    makriTensorPropagatorIterativeII1995}
\begin{widetext}
    \begin{align}
        \mel{s^+_N}{\rho(N\Delta t)}{s^-_N}               & = \sum_{s_0^\pm}\sum_{s_1^\pm}\cdots\sum_{s_{N-1}^\pm}\mel{s_N^\pm}{\mathcal{E}_0(\Delta t)}{s_{N-1}^\pm}\mel{s_{N-1}^\pm}{\mathcal{E}_0(\Delta t)}{s_{N-2}^\pm}\cdots\mel{s_1^\pm}{\mathcal{E}_0(\Delta t)}{s_0^\pm}\mel{s_0^+}{\rho(0)}{s_0^-} F\left[\left\{s^\pm_j\right\}\right],\label{eq:basic_pi} \\
        \text{where }F\left[\left\{s^\pm_j\right\}\right] & = \exp\left(-\frac{1}{\hbar}\sum_{k=0}^N\left(s^+_k - s^-_k\right)\sum_{k'=0}^k\left(\eta_{kk'}s^+_{k'} - \eta^*_{kk'}s^-_{k'}\right)\right).
    \end{align}
\end{widetext}
The bare forward-backward propagator, $\mathcal{E}_0(t) = \exp(-i H_0
    t/\hbar)\otimes\exp(i H_0 t/\hbar)$, is the dynamical map corresponding to the
bare system, $s_j^\pm$ is the state of the system at the $j$th time-point, and
$F\left[\left\{s^\pm_j\right\}\right]$ is the Feynman-Vernon influence
functional~\cite{feynmanTheoryGeneralQuantum1963} along
the path $s^\pm_j$. The influence is calculated in terms of the
$\eta$-coefficients~\cite{makriTensorPropagatorIterativeI1995,
    makriTensorPropagatorIterativeII1995} which are discretizations of the bath
response function~\cite{boseZerocostCorrectionsInfluence2022}. However this
formalism does not allow for a simple inclusion of the jump operators.

In order to incorporate the Lindblad jump operators, we start with the
Nakajima-Zwanzig master equation~\cite{nakajimaQuantumTheoryTransport1958,
    zwanzigEnsembleMethodTheory1960}. The reduced density matrix of the system,
$\rho(t)$, simulated using Eq.~\ref{eq:basic_pi}, also satisfies the
Nakajima-Zwanzig master equation:
\begin{align}
    \dot{\rho}(t) & = -\frac{i}{\hbar}\mathcal{L}_0\rho(t) + \int_{0}^{\tau_\text{mem}} \mathcal{K}(\tau)\rho(t-\tau)\dd{\tau},\label{eq:nakajima_zwanzig}
\end{align}
where $\tau_\text{mem}$ is the length of the non-Markovian memory, and
$\mathcal{L}_0\cdot = \commutator{H_0}{\cdot}$ is the system Liouvillian.
Generally obtaining the memory kernel is a tricky affair. A lot of work has gone
into trying to obtain these memory kernel elements accurately from either
approximate~\cite{mulvihillCombiningMappingHamiltonian2019,
    mulvihillModifiedApproachSimulating2019} or numerically
exact~\cite{chatterjeeRealTimePathIntegral2019} quantum dynamical simulations of
the evolution of the reduced density matrix.

A simpler and more aesthetically pleasing route was proposed by Cerrillo and
Cao~\cite{cerrilloNonMarkovianDynamicalMaps2014}. They showed that from the
dynamical maps that relate the time-evolved density matrix $\rho(t)$ to the
initial density matrix $\rho(0)$ under the influence of the environment,
$\rho(t) = \mathcal{E}(t)\rho(0)$, one can derive the transfer tensors, $T_{l}$
that satisfy
\begin{align}
    \rho(t_n) & = \sum_{k=1}^{L} T_k\,\rho(t_{n-k}).\label{eq:transfer_tensor}
\end{align}
The dynamical maps including the environment effects can be simulated using
Eq.~\ref{eq:basic_pi} by not contracting the initial reduced density matrix. The
transfer tensor method (TTM) has already been used with path integral-based
simulations~\cite{boseQuantumDynamicsJlModular2023,
    boseImpactSpatialInhomogeneity2023}.  It has also been shown that for short
time-steps, the transfer tensors can be related to the memory kernel
by~\cite{cerrilloNonMarkovianDynamicalMaps2014}
\begin{align}
    T_k & = (1 - i\mathcal{L}_0\Delta t)\delta_{k,1} + \mathcal{K}_k\Delta t^2.\label{eq:short_time_kernel}
\end{align}
For longer time steps, this would break down because it is based on a
discretization of the time-derivative of $\rho(t)$ correct to
$\mathcal{O}(\Delta t)$. A better mapping can be obtained by converting to the
reduced density matrix to the interaction picture with respect to the bare
system Hamiltonian, $H_0$, in terms of the bare dynamical map,
$\mathcal{E}_0(t)$,
\begin{align}
    \tilde{\rho}(t)                      & = \mathcal{E}_0^{-1}(t)\rho(t)\label{eq:rho_interaction}                                \\
    \text{where }\mathcal{E}_0(t)\rho(t) & = \exp\left(\frac{-i H_0 t}{\hbar}\right)\rho(t)\exp\left(\frac{i H_0 t}{\hbar}\right).
\end{align}
The equation of motion for the interaction picture reduced density matrix is
\begin{align}
    \dot{\tilde{\rho}}(t) & = \mathcal{E}^{-1}_0(t)\int_{0}^{\tau_\text{mem}} \mathcal{K}(\tau)\,\mathcal{E}_0(t-\tau)\,\tilde{\rho}(t-\tau) \dd{\tau}.
\end{align}
Consequently,
\begin{align}
    \tilde{\rho}(t_n) & = \tilde{\rho}(t_{n-1}) + \left(\Delta t\right)^2\mathcal{E}^{-1}_0(t_n)\sum_{k=1}^L \mathcal{K}_k\,\mathcal{E}_0(t_{n-k})\,\tilde{\rho}(t_{n-k}).\label{eq:GQME_interaction}
\end{align}
Converting Eq.~\ref{eq:transfer_tensor} to interaction picture, the
corresponding equation in terms of the transfer tensors becomes
\begin{align}
    \tilde{\rho}(t_n) & = \mathcal{E}^{-1}_0(t_n)\sum_{k=1}^L T_k\,\mathcal{E}_0(t_{n-k})\,\tilde{\rho}(t_{n-k}).\label{eq:TTM_interaction}
\end{align}

Comparing Eqs.~\ref{eq:GQME_interaction} and~\ref{eq:TTM_interaction}, we obtain
\begin{align}
    T_k & = \mathcal{E}_0(\Delta t)\delta_{k,1} + \mathcal{K}_k\Delta t^2\label{eq:ttm_memorykernel}
\end{align}
as an analogue of
Eq.~\ref{eq:short_time_kernel}~\cite{cerrilloNonMarkovianDynamicalMaps2014}. The
main benefit being that this allows us to take advantage of the larger
time-steps enabled by ideas like the quasi-adiabatic
propagator~\cite{makriNumericalPathIntegral1995}. Earlier work involving
evaluating the memory kernel from numerically exact quantum dynamical
simulations~\cite{chatterjeeRealTimePathIntegral2019} required interpolation of
the path integral results which can bring in errors if not done carefully.

To incorporate the Lindbladian jump operators, we modify the Nakajima-Zwanzig
master equation for the reduced density matrix, Eq.~\ref{eq:nakajima_zwanzig}, as
\begin{align}
    \dot{\rho}^{(L)}(t) & = -\frac{i}{\hbar}\mathcal{L}_0\rho^{(L)}(t) + \int_{0}^{\tau_\text{mem}} \mathcal{K}(\tau)\rho^{(L)}(t-\tau)\dd{\tau}\nonumber \\
                        & + \sum_j \left(L_j\rho^{(L)}(t)L^\dag_j - \frac{1}{2}\anticommutator{L^\dag_jL_j}{\rho^{(L)}(t)}\right),
\end{align}
where the superscript $(L)$ denotes that this equation involves the empirical
terms from the Lindbladian master equation.

Let us, once again, cast the density matrix in an interaction picture
vis-\`a-vis the bare system evolution, Eq.~\ref{eq:rho_interaction}, which in
presence of the jump operators satisfies
\begin{align}
    \dot{\tilde{\rho}}^{(L)}(t) & = \mathcal{E}_0^{-1}(t)\left(\int_{0}^{\tau_\text{mem}} \mathcal{K}(\tau)\rho^{(L)}(t-\tau)\dd{\tau}\right.\nonumber \\
                                & +\left. \sum_j \left(L_j\rho^{(L)}(t)L^\dag_j - \frac{1}{2}\anticommutator{L^\dag_jL_j}{\rho^{(L)}(t)}\right)\right)
\end{align}
Consequently we have the following expression for the time evolution of the
reduced density matrix:
\begin{align}
    \rho^{(L)}_n & = \mathcal{E}_0(\Delta t)\rho^{(L)}_{n-1} + \sum_{j=1}^{L} \mathcal{K}_j \rho^{(L)}_{n-j}\Delta t^2\nonumber                                    \\
                 & + \sum_j \left(L_j\rho^{(L)}_{n-1}L^\dag_j - \frac{1}{2}\anticommutator{L^\dag_jL_j}{\rho^{(L)}_{n-1}}\right) \Delta t,\label{eq:Lindblad_NZME}
\end{align}
which along with Eq.~\ref{eq:ttm_memorykernel} form the final results of this
communication. The transfer tensors may be obtained directly from the augmented
propagators or dynamical maps produced by any path integral
technique~\cite{makriTensorPropagatorIterativeI1995,
    strathearnEfficientNonMarkovianQuantum2018, makriSmallMatrixDisentanglement2020,
    bosePairwiseConnectedTensor2022}.

\begin{figure}
    \includegraphics[scale=0.3]{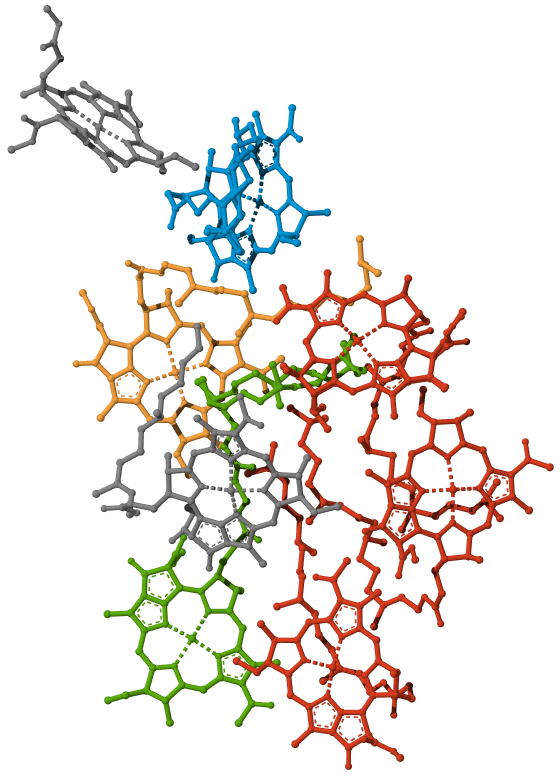}
    \caption{Fenna--Matthews--Olson Complex. Four-site coarse graining as indicated by the colors.}\label{fig:fmo}
\end{figure}

\begin{figure}
    \includegraphics{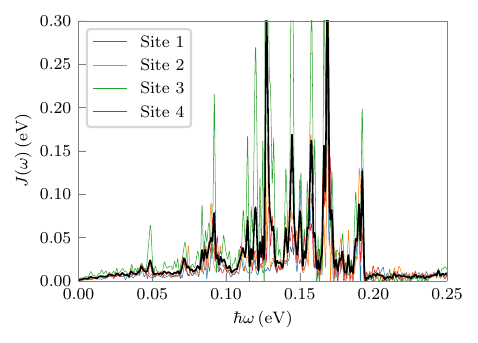}
    \caption{Site-specific and average spectral densities (black solid line), obtained by Maity \textit{et al.}~\cite{maityDFTBMMMolecular2020}, characterizing the chemical environment of the chromophores.}\label{fig:fmo_spectral_density}
\end{figure}

\emph{Numerical Results.} As an example of this methodology, consider the
four-site model of the Fenna-Matthews-Olson complex, Fig.~\ref{fig:fmo},
studied in Ref.~\cite{boseImpactSolventStatetoState2023}. This model is
described by the vibronic Hamiltonian:
\begin{align}
    H            & = H_0 + H_\text{env}                                                                                                                                              \\
    H_0          & = \sum_{j=1}^4 \epsilon_j\dyad{j} + \sum_{j\ne k}h_{jk}(\dyad{j}{k}+\dyad{k}{j})                                                                                  \\
    H_\text{env} & = \sum_{k=1}^4\sum_{j=1}^{N_\text{osc}}\frac{p_{jk}^2}{2m_{jk}} + \frac{1}{2}m_{jk}\omega_{jk}^2\left(x_{jk} - \frac{c_{jk}\dyad{k}}{m_{jk}\omega{jk}^2}\right)^2
\end{align}
where $H_0$ is the electronic ``system'' Hamiltonian and $H_\text{env}$
describes the environment and its interaction with the system. This environment
consisting of molecular nuclear motion and motion of the protein scaffolding, is
mapped to site-specific baths of harmonic oscillators. The frequency and the
coupling of the $j$th bath mode corresponding to the $k$th chromophoric site is
characterized by a spectral density
\begin{align}
    J_k(\omega) & = \frac{\pi}{2}\sum_{j=1}^{N_\text{osc}}\frac{c_{jk}^2}{m_{jk}\omega_{jk}}\delta(\omega-\omega_{jk}),
\end{align}
which can be obtained from classical trajectory-based simulations of bath
correlation functions. The spectral density corresponding to the molecular
vibrations and protein scaffolding was characterized by Maity \textit{et
    al.}~\cite{maityDFTBMMMolecular2020} using QM/MM. The ones corresponding to the
first four sites is shown in Fig.~\ref{fig:fmo_spectral_density}. Also note that
$\ket{j}$ corresponds to the singly excited state where the $j$th molecule is
excited and all the others are in the ground state.

Now, this Hamiltonian describes a Frenkel-Holstein model and therefore, does not
account for the extraction of the exciton at chlorophyll site number 3. We can
account for that empirically either using non-Hermitian Hamiltonians or by using
the Lindblad master equation. The Lindblad master equation has the advantage of
having the ground state population increase corresponding to the extraction of
the exciton. To enable this, we need to expand our Hilbert space by also
considering $\ket{g}$ corresponding to the state in which all the chlorophyll
molecules are in the ground state. In this case there is only one jump operator,
$L = \gamma\dyad{g}{3}$, which is a non-Hermitian lowering operator that
de-excites the 3rd molecule. The timescale in which this effect happens is given
by $\frac{1}{\gamma^2}$.

\begin{figure}
    \includegraphics{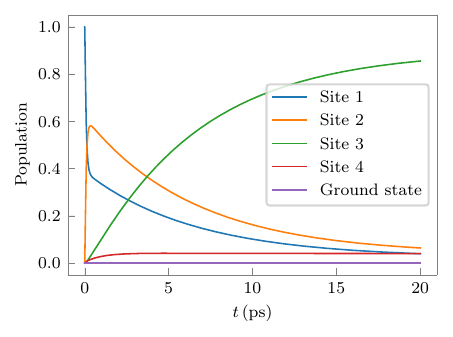}
    \caption{Dynamics of the excitonic population without the Lindbladian jump operators.}\label{fig:base_fmo}
\end{figure}

\begin{figure}
    \hspace*{-0.25cm}
    \subfloat[\SI{2.5}{\ps} decay.]{\includegraphics{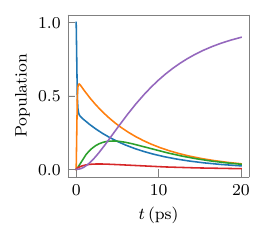}}
    ~\subfloat[\SI{5.0}{\ps} decay.]{\includegraphics{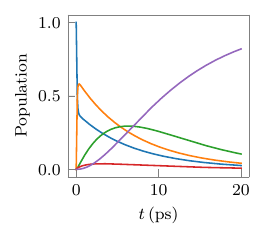}}

    \hspace*{-0.25cm}
    \subfloat[\SI{10.0}{\ps} decay.]{\includegraphics{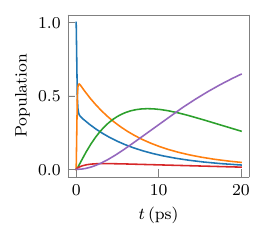}}
    ~\subfloat[\SI{200.0}{\ps} decay.]{\includegraphics{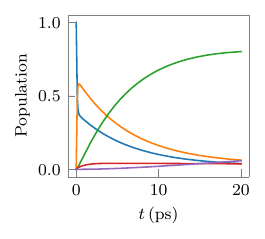}}
    \caption{Dynamics of the excitonic population with different decays on the 3rd chromophore. The colors used are the same as Fig.~\ref{fig:base_fmo}.}\label{fig:fmo_10ps}
\end{figure}

The base dynamics without the Lindblad jump operators was first calculated with
a time-step of $\Delta t=\SI{3}{\fs}$. The dynamical maps, $\mathcal{E}(t)$, was
generated using the time-evolving matrix product state (TEMPO)
algorithm~\cite{strathearnEfficientNonMarkovianQuantum2018,
    boseTensorNetworkRepresentation2021} as implemented in the
\texttt{QuantumDynamics.jl} package~\cite{boseQuantumDynamicsJlModular2023}. The
maps were calculated upto a maximum time of \SI{300}{\fs}, with a memory length
of $\tau_\text{mem} = \SI{150}{\fs}$. The transfer tensors were derived from
this data and used to propagate the dynamics up to a time of \SI{5}{\ps}. The
dynamics corresponding to $\rho(0)=\dyad{1}$ is shown in
Fig.~\ref{fig:base_fmo}. The Frenkel model of exciton transport conserves the
number of excitons. Consequently, when we start with an initial excitation on
$\rho(0)=\dyad{1}$, the dynamics remains in the manifold of singly excited
states. The population of the ground state remains identically zero throughout
the dynamics.

\begin{figure}
    \hspace*{-0.25cm}
    \subfloat[No decay of 3rd site]{\includegraphics{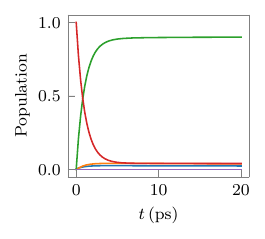}}
    ~\subfloat[\SI{2.5}{\ps} decay.]{\includegraphics{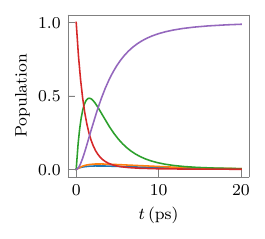}}

    \caption{Comparison of dynamics starting from $\rho(0)=\dyad{4}$ with and without decay from the 3rd site.}\label{fig:decay_from_4}
\end{figure}

Now, we add the Lindbladian jump operator with time-scales of
$\frac{1}{\gamma^2}=\SIlist{2.5;5;10;200}{\ps}$. We solve
Eq.~\ref{eq:Lindblad_NZME} by using the memory kernel obtained from the transfer
tensors in the previous step. The results are shown in Fig.~\ref{fig:fmo_10ps}.
Once the memory kernel has been generated from the path integral dynamical maps,
the incorporation of the Lindbladian jump operators are free. As expected, the
rate of the growth of the population of the ground state, $\ket{g}$, is directly
related to the time-scale of the decay. However, this decay of the 3rd site
leads to subtle changes in the population dynamics of the other sites as well.
These are quantitatively accounted for. The dynamics under a very long decay
time of $\frac{1}{\gamma^2}=\SI{200}{\ps}$, Fig.~\ref{fig:fmo_10ps}~(d), is
shown to demonstrate the asymptotic convergence of the path integral Lindblad
dynamics to the path integral dynamics in absence of any Lindblad jump operators
(Fig.~\ref{fig:base_fmo}). Finally, probably the change due to the decay from
the 3rd site is the greatest if we take an initial condition localized on the
4th site ($\rho(0)=\dyad{4}$). As shown in Fig.~\ref{fig:decay_from_4}, the rise
in the other sites is all but quenched when the decay is brought in.

In this case, once the memory kernels were obtained from the initial path
integral simulations via the transfer tensors, adding any jump operator or
changing the strength of the jump operator amounts to resolving the
Eq.~\ref{eq:Lindblad_NZME} with the corresponding changes. Since this is just a
solution of a difference equation, it is effectively free and can be done with
great efficiency.

\emph{Conclusion.} In this communication, we have developed a method for doing
path integral Lindblad dynamics. The goal is to unify the rigorous path integral
methods for the environment degrees of freedom that can be properly
characterized in terms of spectral densities with the Lindblad master equation
approach of dealing with phenomena that are described only by some empirical
time-scales. This approach brings together the best of both worlds, maximizing
the numerical accuracy of the simulations while not incurring any additional
costs. This path integral Lindblad dynamics method can be used to incorporate a
variety of empirically specified phenomenon like spontaneous emission life-time
of states, leakage from particular states, and others, with rigorous path
integral descriptions of dynamics under the influence of thermal environments
specified by spectral densities. This combination can be done at no extra cost
because the solution to the master equation is free in comparison to the path
integral simulation. Even changing the Lindbladian jump operators to model
different processes can be done extremely efficiently once the memory kernels
have been initially obtained.

While this method is already very useful in studying realistic problems, future
developments would be able to leverage its power even more significantly. We
will extend the path integral Lindblad equations to study the changes brought
about by the presence of thermal environments by phenomena represented by
lindblad jump operators. For example, it would be interesting to understand how
the presence of a leaky cavity changes the absorption spectrum and transfer
pathways of a exciton-polaritonic system. The path integral Lindblad dynamics
method has been implemented in the \texttt{QuantumDynamics.jl}
package~\cite{boseQuantumDynamicsJlModular2023} for simulating the dynamics of
open quantum systems.

\bibliography{library}
\end{document}